\documentclass[12pt]{article}
\usepackage{a4,amssymb,verbatim,amsmath}
\numberwithin{equation}{section}

\def\dx#1{{\partial \over \partial#1}}

\def\dhp#1{\mathop {#1}\limits_{+h}}
\def\dhm#1{ \mathop{#1}\limits_{-h}}

\def\dd#1{ \mathop{#1}\limits_{+h}}

\def\dpm#1{ \mathop{#1}\limits_{\pm h}}

\newcommand{\ddx}{\partial \over \partial x}
\newcommand{\ddy}{\partial \over \partial y}

\begin{document}

\hspace{4cm}

\begin{center}
{\Large {\bf Continuous symmetries of Lagrangians }}\\
\medskip
{\Large {\bf  and exact solutions of discrete equations}} \\
\medskip

\end{center}

\bigskip
\bigskip

\begin{center}
{\large Vladimir Dorodnitsyn}$^{*}$,
{\large Roman Kozlov}$^{\dag}$
{\large and Pavel Winternitz}$^{\ddag}$
\medskip
\hspace{1.5 cm}

${}^{*}$
Keldysh Institute of Applied Mathematics of Russian
Academy of Science,\\
Miusskaya Pl.~4, Moscow, 125047, Russia; \\
E-mail address: dorod@spp.Keldysh.ru \\
${}^{\dag}$   Department of Informatics,
University of Oslo, 0371, Oslo, Norway;  \\
E-mail address: kozlov@ifi.uio.no 

${}^{\ddag}$ Centre de Recherches Math\'ematiques 
et D\'epartement de math\'ematiques \\
et de statistique, Universit\'e de Montr\'eal, \\
Montr\'eal, QC, H3C 3J7, Canada; \\
E-mail address: wintern@crm.umontreal.ca

\bigskip
\bigskip

  { \large July 15, 2003}
  \end{center}

\bigskip
\bigskip
\begin{center}
{\bf Abstract}
\end{center}
\begin {quotation}

One of the difficulties encountered when studying physical theories in
discrete space-time is that of describing the underlying continuous
symmetries (like Lorentz, or Galilei invariance). One of the ways of
addressing this difficulty is to consider point transformations acting
simultaneously on difference equations and lattices. In a previous article
we have classified ordinary difference schemes invariant under Lie groups
of point transformations. The present article is devoted to an invariant
Lagrangian formalism for scalar single-variable difference schemes. The
formalism is used to obtain first integrals and explicit exact solutions
of the schemes. Equations invariant under two- and three- dimensional
groups of Lagrangian symmetries are considered.
\end{quotation}

\bigskip

\eject

\section{\large \bf Introduction}

A recent article was devoted to a symmetry classification of 
second order ordinary difference equations [1]. This  
was modeled on a  paper by Sophus Lie, in which he 
provided a symmetry classification of second order
differential 
equations (ODEs)~\cite{[1]}. As a matter of fact, the
classification 
of difference schemes goes over into Lie's
classification of ODEs 
in the continuous limit~\cite{[45]}. 

S.~Lie showed that a second order ODE can be invariant
under a group 
$G_r$ of dimension $N = 0, 1, 2, 3, \mbox{or}\  8$. 
For $N \geq 2$ the equation can be integrated in
quadratures. 
This can be done by transforming the equation to one
of the ``canonical'' 
forms, integrated by Lie himself~\cite{[1]}. Virtually
all 
standard methods of integrating second order ODEs
analytically can 
be interpreted in this manner (though this is not
mentioned in most 
elementary textbooks). 

The situation with difference equations is much less developed. 
This is not surprising, since applications of Lie
group theory 
to difference equations are much more recent [3,...,27]. Several 
different approaches are being pursued. One
possibility 
is to consider the difference equations on a fixed
lattice [3,...,13] 
and consider only transformations that do not act on the lattice. 
In order to obtain physically interesting symmetries
in this approach, 
it is necessary to go beyond point symmetries and to
let the transformations 
act on more than one point of the lattice. Lie algebra
contractions 
occur in the continuous limit and some ``generalized''
symmetries 
may ``contract'' to point ones~\cite{[10]}.

The second possibility is to consider group
transformations acting 
both on the difference equations and on the lattice
[1,17,...,27]. 
Technically, for systems involving one dependent and 
one independent variable, this is achieved by considering a
difference scheme, 
consisting of two equations, one representing the
actual difference 
equation, the other the lattice. 

This is the approach that we will follow in the
present article. More 
specifically, we will consider the same three--point
scheme as in 
our previous article~\cite{[45]}. The continuous limit
of the scheme, 
if it exists, will be a second order ODE. 

Thus, we consider two variables, $x$ and $y$, with $x$
the independent one 
and $y$ dependent. The variable $x$ runs through an
infinite set of values 
$ \{ x = x_k, \ k \in \mathbb{Z} \}$ that are not
necessarily 
equally spaced and are not prescribed a priori.
Instead, we 
give two relations between any three neighboring points
\begin{equation} \label{sysa1}
F ( x,x_{-},x_{+},y,y_{-},y_{+} ) = 0, 
\end{equation} 
\begin{equation} \label{sysa2}
\Omega (x,x_{-},x_{+},y,y_{-},y_{+}) = 0
\end{equation} 
and also specify some initial conditions like $x_0$, 
$x_1$, 
$ y_0 = y (x_0)$,  $ y_1 = y (x_1)$. In the continuous
limit 
eq.~(\ref{sysa1}) goes into an ODE, (\ref{sysa2}) into 
an identity (like $0 = 0 $), if the continuous limit
exists. 
\par
The group transformations considered in this approach
are of the 
same type as for ODEs. They are generated by a Lie
algebra 
of vector fields of the form 
\begin{equation} \label{operator1}
 X =   \xi (x,y) { \ddx } + \eta (x,y) {\ddy }. 
\end{equation}
The corresponding transformations are purely point
ones, since 
the coefficients $ \xi$ and $\eta$ depend only on
$(x,y)$, not 
on the shifted points  $(x_{+},y_{+})$ or
$(x_{-},y_{-})$. 

In ~\cite{[45]} we showed how Lie group theory can
be used to classify 
such pairs of equations as (\ref{sysa1}) and
(\ref{sysa2}). Possible 
dimensions of the symmetry group $G$ of 
eq.~(\ref{sysa1}), (\ref{sysa2}) are  
$N = 0, 1,  2, 3, 4, 5,   \mbox{and}\  6$. The highest
dimension, 
$N = 6$, occurs only for difference schemes equivalent 
to 
$$
{ y_{+} - 2y + y_{-} \over ( x_{+} -x )^2 } = 0 ,
\qquad 
x_{+} - 2x + x_{-} = 0 . 
$$
The purpose of this article is to provide a Lagrange
formalism 
and difference analog of Noether's theorem for 
second order difference schemes of the form 
(\ref{sysa1}) and (\ref{sysa2}), 
admitting Lie point symmetry groups. 
The Lagrangians will be used to obtain first integrals
and exact analytic solutions of
the difference schemes.  
  
\medskip

\section{\large \bf General theory}
\medskip

\subsection{Definitions and notations} 

We study the difference system (\ref{sysa1}) and
(\ref{sysa2}). 
In general, we assume that these equations can be
solved 
to express $x_{+}$ and $y_{+}$ explicitly in terms of 
$( x, y, x_{-}, y_{-} ) $ and also vice verse, i.e.
$(x_{-}, y_{-}) $ in
terms of the other quantities. We also make use of the
following 
quantities 
\begin{equation} \label{formur}
\begin{array}{c}
{ \displaystyle 
h_{+} = x_{+} - x, \qquad  h_{-} = x - x_{-},  \qquad
y_{x} = { y_{+} -y \over h_{+} }, \qquad
y_{\bar{x}} = { y - y_{-}  \over h_{-} } } , \\
\\
{ \displaystyle 
y_{x\bar{x}} = { 2 \over h_{+} + h_{-} } ( y_{x} -
y_{\bar{x}} ), } \\
\end{array}
\end{equation}
i.e. the up and down spacings in $x$, the right and
left discrete first 
derivatives and the discrete second derivative,
respectively. It is also convenient to use the following total shift and discrete differentiation operators:
$$
{ \dpm S} f(x) = f ( x_{\pm} ) ,    
\quad 
{ \displaystyle 
{ \dpm D } = { {  \dpm S } -1 \over \pm h_{ \pm} } } . 
$$
Continuous first and second derivatives are denoted
$y'$ and $y''$, 
respectively. 

When acting on differential equations, the vector
fields (\ref{operator1}) 
must be prolonged to act on derivatives. For
difference schemes, 
the prolongation of a vector field acts on variables 
at other points of the lattice. It is obtained by 
shifting the coefficients to the corresponding points.
For 
three point schemes we have 
\begin{equation} \label{operyx}
\begin{array}{c}
{ \displaystyle {\bf pr}   X  = 
 X  +
 \xi (x_{-},y_{-})  {\partial \over \partial x_{-}  }
+
 \xi (x_{+},y_{+})  {\partial \over \partial x_{+}  } 
}
{ \displaystyle
+  \eta (x_{-},y_{-})  {\partial \over \partial y_{-} 
} +
 \eta (x_{+},y_{+})  {\partial \over \partial y_{+} 
}.  } \\
\end{array} 
\end{equation}

\medskip

\subsection{\large \bf Lagrangian formulation for 
 second order ODEs}

It has been known since E.~Noether's fundamental work
that 
conservation laws for differential equations are
connected with 
their symmetry properties [28,...,31]. For
convenience 
we present here some well--known results adapted to
the case 
of second order ODEs. 

Let us consider the functional 
\begin{equation}   \label{lagyr}
\mathbb{L}   (y) =  \int_{I} L ( x,y,y') dx , \ \ \
I  \subset   \mathbb{R}^1 , 
\end{equation}
where $L ( x,y,y')$ is called a first order
Lagrangian. The functional 
(\ref{lagyr}) achieves its extremal values when $y(x)$
satisfies 
the  Euler-Lagrange equation
\begin{equation}   \label{extyr}
{ \delta L \over \delta y } =
{\partial L \over \partial y  }
- D \left( {\partial L \over \partial y' } \right) =
0, 
\qquad 
D = {\ddx} + y' { \ddy} + y'' { \partial \over
\partial y' } + \cdots , 
\end{equation}
where $D$ is the total derivative operator. 
The equation (\ref{extyr}) is an ODE that can be
rewritten as
\begin{equation}  \label{yyy}
 y'' = f(x,y,y').
\end{equation}

Let us consider a Lie point transformation $G$
generated 
by the vector field  (\ref{operator1}). The group $G$
is a 
``variational symmetry'' of the functional $  { \cal 
L} $ if and  only 
if the Lagrangian satisfies 
\begin{equation} \label{cond}
\mbox{pr}  X(L) + L D( \xi) = 0 ,
\end{equation}
when $\mbox{pr} X$ is the first prolongation of the vector field $X$ for $y'$.
We will actually need a weaker invariance condition
than given by eq.~(\ref{cond}). The vector field $X$ is an 
``infinitesimal divergence symmetry'' of the
functional ${\cal L } (y) $ 
if there exists a function $V(x,y)$ such that [28]
\begin{equation} \label{cong}
\mbox{pr} X(L) + L D( \xi) = D(V), \qquad  V = V(x,y).
\end{equation}

\medskip

\noindent The two important statements for us are:

{\bf 1.} If $X$ is an infinitesimal divergence
symmetry of the functional 
${\cal L } $, it generates a symmetry group of the
corresponding 
Euler-Lagrange equation. The symmetry group of eq. (\ref{extyr}) 
can of course be larger than
the one generated by symmetries of the Lagrangian.

{\bf 2.} Noether's theorem [28,...,31] can be based  on the following
Noether-type identity [31], which holds for any vector field 
and any function $L$:
\begin{equation} \label{ident}
\mbox{pr}  X(L) + L D( \xi) = (\eta - \xi y'){ \delta L \over \delta y } +
D(\xi L + (\eta - \xi y') {\partial L \over \partial y' }).
\end{equation}

It follows that if $X$ is a divergence symmetry of $L$, i.e. 
(\ref{cond}) or (\ref{cong}) is satisfied,
then there exists a first integral

\begin{equation} \label{cdrtre}
\xi L + (\eta - \xi y') {\partial L \over \partial y' } - V = K =const
\end{equation}
 of the corresponding Euler-Lagrange equation. 
\par
The above considerations tell us how to obtain
invariant ODEs
and conservation laws from divergence invariant
Lagrangians. 
They do not tell us how to obtain invariant
Lagrangians for 
invariant equations. This amounts to ``variational
integration'', 
as opposed to variational differentiation. 

\medskip

\noindent A procedure that we shall use below to find
invariant Lagrangians for differential equations
can be summed up as follows.

Start from a given ODE $ y'' = f ( x,y,y') $
and its symmetry 
algebra with basis 
$$ 
X_{\alpha} = \xi_{\alpha} (x,y) {\ddx} + 
\eta_{\alpha} (x,y) {\ddy}, 
\qquad 
\alpha = 1, ..., k. 
$$
Find the invariants of $X_{\alpha}$ in the space $ \{
x,y,y',  \Lambda \} $, 
where $\Lambda $ is the  Lagrangian. 
The appropriate prolongation in this case is 
\begin{equation} \label{prol1}
{ pr }  X = \xi { \ddx} + \eta { \ddy} + 
\zeta^1 { \partial \over  \partial y' } - 
( D \xi )   \Lambda  { \partial \over  \partial 
\Lambda }, 
\qquad  
\zeta ^1 = D(  \eta ) - y' D(  \xi  ) 
\end{equation}
and we require that $ L  (x,y,y') $ should satisfy
\begin{equation} \label{prol2}
{ pr } X (  \Lambda -  { L } ) |_{ \Lambda =  L  }
= 0 .
\end{equation}
Each basis element $X_{\alpha}$ provides us with an
equation 
of the form 
\begin{equation} \label{prol3}
 \xi_{\alpha}  { \partial L \over \partial x } 
+ \eta_{\alpha} { \partial L \over \partial y } 
+ \zeta_{\alpha}^1 { \partial L \over \partial y' } -
L D (\xi_{\alpha} ) = 0 .
\end{equation}

Solve the partial differential equations
(\ref{prol3}). This will 
give us the general form of an invariant
Lagrangian. It may 
involve arbitrary functions of the invariants of $X$.

Request that the Euler-Lagrange equation (\ref{extyr}) should 
coincide with the equation we started from. This will further 
restrict the invariant Lagrangian and determine whether one exists.

If this procedure does not yield a suitable Lagrangian,
then step~1 can be weakened. We can request that the Lagrangian be
invariant under some subgroup of the symmetry group of the given ODE, rather 
then the entire group. We then go through step~2, then verify 
whether the obtained Lagrangian is divergence invariant under 
the entire group, or at least  a larger subgroup. In
any case, each divergence symmetry of the Lagrangian will provide a 
first integral of the ODE.    

\par
For  ODEs the Lagrangian formalism is not the only integration method.
The existence of one-parameter symmetry group provides a reduction to a first-order ODE
directly. The existence of a two-parameter symmetry group makes it possible to integrate in quadratures.
 An invariant Lagrangian provides an  alternative. 
Indeed, assume that we know two first integrals
\begin{equation} \label{wow}
f_1 (x,y,y') = A, \qquad f_2 (x,y,y') = B,
\end{equation}
then we eliminate $y'$ from these two equations and obtain the general solution

\begin{equation} \label{vova}
y = F(x,A,B),
\end{equation}
of the corresponding ODE (\ref{yyy}) by purely {\it algebraic} manipulations. It is this method
of invariant Lagrangians that generalizes to difference equations and is particularly
useful when direct methods fail.

\medskip

\subsection{\large \bf Lagrangian formalism for second order difference equations}

The variational formulation of discrete equations and
a discrete analog of 
Noether's theorem are much more recent [19,25,26,27].
Here we briefly  overview the  results that we shall
need below.

 Let us consider 
a finite difference functional
\begin{equation} \label{in1}
\mathbb{L}_{h} = \sum^{}_{\Omega} {\cal L}(x,x_{+},y,y_{+}) h_{+} ,
\end{equation}
defined on some one--dimensional lattice ${\Omega}$
with 
step $h_{+}$ that generally can depend on the solution  
\begin{equation} \label{in2}
h_{+} = \varphi(x,y,x_{+}, y_{+}).
\end{equation}

The Lagrangian (\ref{in1}) must be considered together with a lattice (2.16). On different
lattices it can have different continuous limits and in this limit the lattice
equation itself vanishes (turns into an identity like $0=0$)
\par
In the continuous case, a Lagrangian $L$ provides an equation (the Euler-Lagrange equation)
that inherits all the symmetries of $L$. In the discrete case we wish the Lagrangian (\ref{in1})
to provide two equations: the entire difference system (1.1),(1.2). 
Moreover, the three-point difference system should 
inherit the symmetries of two-point Lagrangian.

\par
Let us again consider a Lie group of point 
transformations, generated by a Lie algebra of vector
fields 
$ X_{ \alpha } $ of the form (\ref{operator1}). 
The infinitesimal invariance condition of the
functional (\ref{in1}) 
on the lattice (\ref{in2}) is given by two equations [19,25,27]:
\begin{equation} \label{inter1}
\begin{array}{c} 
{ \displaystyle 
\xi        \frac{\partial {\cal L}}{\partial x} +
\xi^+ \frac{\partial {\cal L}}{\partial {x_{+} }} +
\eta \frac{\partial {\cal L}}{\partial y} +
\eta^+  \frac{\partial {\cal L}}{\partial {y_{+} }} + 
{\cal L} { \dhp D }
(\xi  ) = 0, } \\
\\
{ \displaystyle 
\dhp S(\xi  ) - \xi   = X  (\varphi  ), } \\
\end{array} 
\end{equation}
where 
\begin{equation} \label{inter2}
\begin{array}{c} 
{ \displaystyle \xi^+ = \xi ( x_{+}, y_{+} ) , 
\quad 
\eta^+ = \eta  ( x_{+}, y_{+} ). 
 } \\
\end{array} 
\end{equation}

\par

Let us consider a variation of the difference functional (\ref{in1}) along
some curve $ y= \phi (x)$ at some point $(x,y)$. The variation will 
effect only two terms in the sum (\ref{in1}):
\begin{equation} \label{i1}
\mathbb{L}_{h} = ... + {\cal L}(x,x_{-},y,y_{-}) h_{-}+ 
{\cal L}(x,x_{+},y,y_{+}) h_{+} + ... ,
\end{equation}
so we get the following expression for the variation of the difference
functional
\begin{equation} \label{i5}
\delta L =  \frac{\delta{\cal L}}{\delta x}  \delta x
+   \frac{\delta{\cal L}}{\delta y} \delta y,
\end{equation}
where $\delta y = {\phi}' \delta x$ and
$$
\frac{\delta{\cal L}}{\delta x}=
h_{+} \frac{\partial {\cal L}}{\partial x} +
h_{-} { \frac{\partial {\cal L}}{\partial x} }^{-}
+{\cal L}^- - {\cal L} , \qquad 
\frac {\delta {\cal L}}{\delta y}
=h_{+} \frac {\partial {\cal L}}{\partial y}
+ h_{-} { \frac {\partial {\cal L}}{\partial y} }^{-}, 
$$
where ${\cal L}^- = \dhm S( {\cal L} )$.

Thus, for an arbitrary curve the stationary value of difference functional
is given by any solution of the {\it two} equations, called  
{\it quasiextremal equations} 
\begin{equation} \label{d55}
  \frac{\delta{\cal L}}{\delta x}  =0, \qquad
   \frac{\delta{\cal L}}{\delta y} =0.
\end{equation}
Both of them tend to the differential Euler-Lagrange equation
in the continuous limit. Together they represent
the entire difference scheme and could be called "the discrete 
Euler-Lagrange system". The difference between these two equations, 
or some other function of them that vanishes in the continuous limit will
represent the lattice.

\par
Now let us consider a vector field (\ref{operator1}) 
with given coefficients
$\xi(x,y)$  and  $\eta(x,y)$. 
Variations along the integral curves of this vector field are given 
by $\delta x = 
\xi da$ and $\delta y = \eta da$, where $da$ is a variation of a group 
parameter.
A   stationary  value of the
difference functional  (\ref{in1}) along the flow generated by this 
vector field is given by the equation:
\begin{equation} \label{in5}
\xi  \frac{\delta{\cal L}}{\delta x}
+ \eta  \frac{\delta{\cal L}}{\delta y} = 0,
\end{equation}
which depends explicitly on the coefficients of the generator .
\par
If we have a Lie algebra of vector fields of dimension 2 or more,
then  a  stationary  value of the
difference functional  (\ref{in1}) along the entire flow will be 
achieved on the intersection of the solutions of all equations of the type 
(\ref{in5}), i.e. on the quasiextremals (\ref{d55}).

\par
On the other hand, 
equations (\ref{d55}) can be interpreted as a three-point difference 
scheme of the form (\ref{sysa1}),(\ref{sysa2}). For instance, given two points $(x,y)$ and
$(x_{-}, y_{-})$, we can calculate $(x_{+}, y_{+})$.
In the continuous limit both of these equations will provide 
the same second-order differential equation. Thus, one of the quasiextremal 
equations can be identified with eq.~(\ref{sysa1}) and 
the difference between the two of them with the lattice equation 
(\ref{sysa2}).

\par
It has been shown elsewhere [19,25,27], that if the
functional  (\ref{in1}) is invariant under some group $G$, 
then the quasiextremal  equations (\ref{d55}) are also invariant 
with respect to $G$. As in the continuous case, 
the quasiextremal equations can be invariant with respect to a larger group 
than the corresponding Lagrangian.

\par
A useful operator identity, valid for any Lagrangian 
$ { \cal L } ( x , x_{+} , y , y_{+} ) $ and any
vector field $X$  is ([19,25]): 

\begin{equation} \label{in38}
\begin{array}{c}
{ \displaystyle
\xi  \frac{\partial {\cal L}}{\partial x} +
\xi^+ \frac{\partial {\cal L}}{\partial {x_{+} }} +
\eta  \frac{\partial {\cal L}}{\partial y} +
\eta^+  \frac{\partial {\cal L}}{\partial {y_{+} }} 
+ {\cal L} { \dhp D } (\xi ) = } \\
\\
{ \displaystyle
= \xi  \left ( \frac{\partial {\cal L}}{\partial x}
+ \frac{h_{-} }{h_{+} } { \frac{\partial {{\cal
L}}}{\partial x} }^{-} -
{\dhp D }({\cal L}^-) \right ) + 
\eta 
\left ( \frac{\partial {\cal L}}{\partial y} +
\frac{h_{-} }{h_{+} }
{ \frac{\partial {{\cal L}}}{\partial y} }^{-} \right
) + } \\
\\
{ \displaystyle 
+ { \dhp D } \left (h_{-} \eta  
{ \frac{\partial {{\cal L}}}{\partial y} }^{-} +
h_{-} \xi  { \frac{\partial {{\cal L}}}{\partial x}
}^{-}
+ \xi  {\cal L}^- \right ) } . 
\end{array}
\end{equation}

\noindent 
From eq. (\ref{in38}) we obtain the following discrete 
analog of Noether's theorem.

\medskip

\noindent {\bf Theorem~2.1} Let the
Lagrangian density ${\cal L}$ be divergence invariant under 
a Lie group $G$ of local point transformations generated by vector fields 
$X$ of the form (\ref{operator1}), i.e. let us have
\begin{equation} \label{in32}
 {\bf pr}    X ( { \cal L } ) +  { \cal L } { \dhp D } ( \xi ) = {
\dhp D } ( V )  
\end{equation}
for some function $V(x,y)$. Then each element $X$ of the Lie algebra 
corresponding to $G$ provides us with a first integral of the 
quasiextremal equations (\ref{d55}), namely 
\begin{equation} \label{in70}
K = h_{-} \eta  
\frac{\partial {{\cal L}}}{\partial y}^- +
h_{-} \xi
\frac{\partial {{\cal L}}}{\partial x}^- + 
\xi {\cal L}^- 
-V  . 
\end{equation}

\noindent {\it Proof.} [19,25] On solutions of the quasiextremal equations
 (\ref{d55}) 
eq.~(\ref{in38}) reduces to 
\begin{equation}
{\dhp D } \left( 
h_{-} \eta  
\frac{\partial {{\cal L}}}{\partial y}^- +
h_{-} \xi
\frac{\partial {{\cal L}}}{\partial x}^- + 
\xi {\cal L}^- 
\right) = 
{ \dhp D } ( V )
\end{equation} 
(we have used eq.~(\ref{in38})). The result (\ref{in70}) follows 
immediately. 
\hfill $\Box$

\medskip

\noindent The fundamental equation (\ref{in70}) is the discrete analog of \
eq.~(\ref{cdrtre}) for ODEs.

\medskip
\par
Let us compare the situation for second order ODEs 
and for 
three-point difference schemes. For a second 
order ODE a Lagrangian 
that is divergence invariant under a two-dimensional
symmetry group provides 
two integrals of motion. From them 
we can eliminate the remaining first derivative and 
obtain the general solution, depending on two
arbitrary constants 
(the two first integrals). Moreover, we do not really
need a Lagrangian. 
Once we have a two dimensional symmetry group of the ODE, we
can integrate in quadratures. 

\par
For three-point difference schemes we have two equations to solve, namely
the system (1.1),(1.2). Equivalently, we have a set of points $(x_n, y_n)$,
labeled by an integer $n$. Any 3 neighboring points are related by two equations that we can write e.g. as
\begin{equation} \label{gauss}
y_{n+1} = F_1 (x_n, y_n, x_{n-1}, y_{n-1} ), \qquad
x_{n+1} = \Omega _1 (x_n, y_n, x_{n-1}, y_{n-1} ).
\end{equation}
Alternatively, the system could be solved for
 $x_{n-1}, y_{n-1}$.  
We mention that we use  notations like 
$x_{n-1} = x_-$, $x_n = x$, $x_{n+1} = x_+$, 
$y_{n-1} = y_-$, $y_n = y$, $y_{n+1} = y_+$   interchangeably.

\par
Given some starting values 
$(x_0, y_0, x_{-1}, y_{-1} )$, we can solve (\ref{gauss}) for $(x_n, y_n )$
with $n \geq 1$, and  $n \leq -2$. The solution will depend on four constants $K_i,
i= 1,...,4$, and can   be written as
\begin{equation} \label{pavel}
y_{n} = y_n ( x_{n}, K_1, K_2, K_3, K_4 ), 
\end{equation}
\begin{equation} \label{pavel1}
x_{n} = x_n ( K_1, K_2, K_3, K_4 ).
\end{equation}
The two quasiextremal equations (\ref{d55})
correspond to the system (\ref{gauss}).
\par
A one-parameter symmetry group of the Lagrangian ${\cal L}$ will
provide us 
with a first integral (\ref{in70}), 
i.e. an equation of the form
\begin{equation} \label{jopa}
f (x_n, y_n, x_{n+1}, y_{n+1} ) = K_1.
\end{equation}
compatible with the system (\ref{d55}). We can solve (\ref{jopa}) 
for e.g. $y_{n+1}$,
substitute into (\ref{gauss}) and thus simplify this system.
\par
A two-dimensional symmetry group will provide two first integrals 
of the form (\ref{in70}). We can solve  for $x_{n+1}$ and $ y_{n+1} $. 
Then system (\ref{gauss}) is reduced to a
 two-point difference scheme. Quite often it is possible to solve it
 by  
integration methods that allow one to integrate a two-point
difference scheme explicitly.
\par
A three-dimensional symmetry group provides three first integrals 
of the type (\ref{in70}). From them we can express 
 $x_{n-1}, y_{n-1}$ and   $ y_{n}$ in terms of  $x_{n}$. This provides us with the solution (\ref{pavel}) and a two-point difference equation relating 
 $x_{n+1}$ and    $x_{n}$. If this equation can be solved, we have a complete solution of the problem. Finally, if we have four first integrals, then we get the general solution of the system by purely algebraic manipulations.

\par
An alternative method can be proposed when
the Lagrangian is invariant with respect to a two-dimensional Lie group.
The discrete Lagrangian corresponding to a given continuous one is not unique and it is possible to introduce a family of Lagrangians:
\begin{equation} \label{lag}
{\cal L}_i = {\cal L}_i (x,x_+, y, y_+, \alpha_i,\beta_i), \quad i =1,2,3,...
\end{equation}
depending on parameters  $\alpha_i,\beta_i$, all satisfying
$$
{\lim}_{(x_+, y_+ ) \to (x,y)}
{\cal L}_i (x,x_+, y, y_+, \alpha_i,\beta_i) = {\cal L} (x,y,y')
$$
for the same continuous Lagrangian ${\cal L} (x,y,y')$. 

\par 
Let us take three different Lagrangians in the family (\ref{lag}),
corresponding to constants $\alpha_1,\beta_1 , \alpha_2,\beta_2$ and
$\alpha_3,\beta_3$. Each of them will lead to two first integrals and
two quasiextremals. In examples considered below we will show that it
is possible to fine-tune the constants  $\alpha_i,\beta_i$ in such a
manner as to get 
a system of two invariant equations of the form
(\ref{sysa1}),(\ref{sysa2}) and three first integrals, yielding a set
of solutions to the two quasiextremal equations. It is these two equations that will constitute the invariant difference system.

In Section 2.2 we described a procedure for obtaining invariant Lagrangians for given
second-order differential equation. For difference equations our
starting point will be a discretization of the continuous
Lagrangian. This is obviously not unique and we shall make use of the
inherent arbitrariness. Once an invariant difference Lagrangian with a
correct continuous limit is chosen we construct the invariant difference scheme in the manner described above.
\par
In our previous article [1] we gave a classification of difference schemes and used all realizations of Lie algebras that provide such schemes. 
Any algebra containing a two-dimensional subalgebra realized by linearly connected vector fields such as
$$
\left( {\ddx} , \quad y{\ddx} \right), \quad 
\left( {\ddx} , \quad x{\ddx} \right),
$$
leads to a linear differential equation and its discretization.
\par
Below we shall consider only genuinely nonlinear difference schemes 
presented in Ref~\cite{[45]} 
that have nonlinear differential equations as their limit.

\medskip

\section{\large \bf Equations corresponding to Lagrangians 
invariant  under
one and two-dimensional groups}

\medskip

\subsection{\large \bf One-dimensional symmetry group}

We start with  the
simplest case of a symmetry group, namely
a one--dimensional group. Its Lie algebra is
generated
by one vector field of the form (\ref{operator1}). By
an
appropriate change of variables we take this vector
field into its rectified form. Thus we have
\begin{equation} \label{op1}
{ \bf D_{1,1}} : \qquad
X_{1}= {\ddy}.
\end{equation}
The most general second order ODE invariant under
$X_1$ is
\begin{equation} \label{eq1}
y''= F( x,y'),
\end{equation}
where $F$ is an arbitrary given function.

Eq.~(\ref{eq1}) is actually already reduced to a first order equation 
for $u = y'$. If $X_1$ is a variational symmetry of eq.~(\ref{eq1}) and 
we know the Lagrangian $L$ that it comes from, we can do better. 
An invariant Lagrangian density will by necessity have the form 
$L = L(x,y')$ (see eq.~(\ref{cond})). The Euler-Lagrange equation 
(\ref{extyr}) reduces to 
\begin{equation} \label{eq101}
{ \partial^2 L \over \partial x \partial y' } + 
y'' { \partial^2 L \over \partial {y'} ^2 }  = 0 
\end{equation}

Substituting for $y''$ from eq.~(\ref{eq1}), we obtain a linear 
partial differential equation for $ L(x,y')$. This of course has an 
infinity of solutions. Let us assume that we know a solution 
$L(x,y) $ explicitly. Eq.~(\ref{cdrtre}), i.e. Noether's theorem, 
provides us with a first integral 
\begin{equation} \label{eq102}
{ \partial L \over  \partial y' } (x,y') =  K  .
\end{equation}
We can solve (in principle) eq.~(\ref{eq102}) for $y'$ as a function of $x$ 
(and $K$). The general solution is then obtained by a quadrature:
\begin{equation} \label{eq103}
y' = \phi (x,K) , \qquad y(x) = y_0 + \int _0 ^x  \phi (x,K) dt  . 
\end{equation}

In the discrete case the situation is similar. Let us assume that 
we know a Lagrangian ${ \cal L}  (x,x_+, y, y_+)$, invariant under the group 
of transformations of $y$, generated by $X_1$ of eq.~(\ref{op1}). 
It will have the form 
\begin{equation} \label{eq104}
{ \cal L} = { \cal L}  (x,x_+, y_x), 
\qquad y_x = { y_+ - y \over x_+ - x} . 
\end{equation}
The corresponding quasiextremal equations, to be identified with 
the system 
(\ref{sysa1}) and (\ref{sysa2}), are 

\begin{equation} \label{ee11a}
{\displaystyle
{ \delta {\cal L} \over \delta y} = } 
{\displaystyle   
- { \partial {\cal L } \over  \partial y_x }  (  x , x_{+} , y_x )  + 
{\partial {\cal L} \over y_{\bar{x}} } (  x_{-} ,  x , y_{ \bar x} ) = 0   } ; 
\end{equation}
\begin{equation} \label{ee11b}
\begin{array}{rl}
{\displaystyle
{ \delta {\cal L} \over \delta x} } & =    
{\displaystyle  
  h_{+} { \partial {\cal L} \over \partial x  } ( x , x_{+} , y_x ) +
y_x { \partial {\cal L} \over \partial y_x  } ( x , x_{+} , y_x )   - 
  {\cal L } ( x , x_{+} , y_x ) } \\
\\
& {\displaystyle 
 +  h_{-} { \partial {\cal L} \over \partial x  } 
(  x_{-} , x , y_{ \bar x} )
-  y_{ \bar x} { \partial {\cal L} \over \partial  y_{ \bar x}   } 
(  x_{-} , x , y_{ \bar x} ) 
+  {\cal L } ( x_{-} ,  x  , y_{ \bar x} ) = 0    } .  
\end{array}
\end{equation}

\noindent The first integral (\ref{in70}) can be read off 
form eq.~(\ref{ee11a}) and is 
\begin{equation} \label{ee11c}
{ \partial {\cal L } \over  \partial y_x }  (  x , x_{+} , y_x )  = K .
\end{equation}
We can solve eq.~(\ref{ee11c}) for $y_x$ and by down shifting obtain 
$y_{\bar{x}}$: 
\begin{equation} \label{ee11d}
y_x = \phi ( x,x_+, K) , \qquad 
y_{\bar{x}} = \phi ( x_{-},x, K). 
\end{equation}
Substituting into the quasiextremal equation (\ref{ee11b}),  
we obtain a relation between $x_+$, $x_-$ and $x$, i.e. a single
 three-point 
relation for the variable $x$. For $y$ we then obtain a two point equation 
\begin{equation} \label{ee11e}
y_+ - y = (x_+ - x) \phi (x, x_+, K) . 
\end{equation}
Equation (\ref{ee11e}) is really a discrete quadrature: a first order 
inhomogeneous linear equation for $y$. 

\bigskip

\noindent {\bf Example~3.1} Consider the Lagrangian 
\begin{equation} \label{ee11f}
{\cal L } = x_n^a x_{n+1}^b \exp (y_x) .
\end{equation}

The quasiextremal equations are 
\begin{equation} \label{ee11g}
- x_n^a x_{n+1}^b \exp (y_x)  + x_{n-1}^a x_{n}^b \exp (y_{\bar{x}})  =0; 
\end{equation}
\begin{equation} \label{ee11h}
\begin{array}{l}
h_+ a  x_n^{a-1} x_{n+1}^b \exp (y_x)  
+ y_x   x_n^{a} x_{n+1}^b \exp (y_x)
- x_n^a x_{n+1}^b \exp (y_x) \\
\\
+ h_- b  x_{n-1}^a x_{n}^{b-1} \exp (y_{\bar{x}}) 
- y_{\bar{x}} x_{n-1}^a x_{n}^{b} \exp (y_{\bar{x}})
+ x_{n-1}^a x_{n}^b \exp (y_{\bar{x}}) =0 .
\end{array}
\end{equation}
The first integral is 
\begin{equation} \label{ee11k}
 x_n^a x_{n+1}^b \exp (y_x)  = K.
\end{equation}
From (\ref{ee11k}) we have 
\begin{equation} \label{ee11m}
y_x = \ln ( K x_n^{-a} x_{n+1}^{-b} ), \qquad 
y_{\bar{x}} = \ln ( K x_{n-1}^{-a} x_{n}^{-b} ).
\end{equation}
Eq.~(\ref{ee11g}) is satisfied identically. Eq.~(\ref{ee11h}) reduces 
to a three point equation for $x$:
\begin{equation} \label{ee11n}
a { x_{n+1} }  + ( b - a) x_n  - b x_{n-1}  + 
x_n ( - b \ln (x_{n+1} ) + ( b - a) \ln (x_{n} ) + a \ln (x_{n-1} ) )    = 0.
\end{equation}
This lattice equation can be reduced to a two-point equation for a new
variable $ \lambda _n = x_{n+1} / x_{n} $: 
\begin{equation} \label{ee11o}
a \lambda _n + ( b-a) - {b \over \lambda _{n-1}} = 
a \ln (  \lambda _{n-1} ) + b \ln (  \lambda _{n} ) . 
\end{equation}
In particular, one can choose the solution 
$\lambda _{n} =  \lambda_{n +1} = \lambda$, where 
$\lambda$ satisfies the equation
$$
a \lambda  + ( b-a) - {b \over \lambda } = 
(a +b)  \ln (  \lambda  ) .
$$
It provides us with the lattice $ x_n = x_0 \lambda ^n$.

Substituting the lattice into (\ref{ee11k}),  
we obtain a two point equation for $y$:
\begin{equation} \label{ee11p}
y_{n+1} - y_{n} = (x_{n+1} - x_n) \ln ( K x_{n}^{-a} x_{n+1} ^{-b} ). 
\end{equation}

The fact that we could solve eq. (\ref{ee11n}) explicitly is specific
for the considered example. The fact that we obtain a three-point
equation involving only the independent variables is true in general.

\medskip

\subsection{\large \bf Two-dimensional symmetry groups}

\noindent ${ \bf D_{2,1} }$ The Abelian Lie algebra
with
non-connected basis elements 
\begin{equation} \label{op21}
X_1 = { \ddx} , \quad X_2 = { \ddy}  
\end{equation}
corresponds to the invariant ODE 
\begin{equation} \label{eq21a}
y''= F(y'),
\end{equation}
where $F$ is an arbitrary function.  

The equation can be obtained from the Lagrangian
\begin{equation} \label{lag21}
L = y + G ( y' ) , \qquad F(y') = { 1 \over G''( y ' ) }.
\end{equation}

The Lagrangian admits symmetries $X_{1}$ and $X_{2}$: 
$$
\begin{array}{l}
\mbox{pr} X_{1} L + L D( {\xi}_{1} ) = 0; \\
\\
\mbox{pr} X_{2} L + L D( {\xi}_{2} ) = 1 = D(x) .\\
\end{array}
$$
With the help of Noether's theorem we obtain the
following first integrals: 
\begin{equation} \label{ddd1}
J_{1}=   y + G( y') - y' G'( y') , \qquad J_{2}= G'(
y') - x  .
\end{equation}
As we mentioned in the Section 2.2, 
it is sufficient to have two first integrals to 
write out the general solution of a second order ODE 
without quadratures. More explicitly, we can solve the second
equation (\ref{ddd1}) for $y'$ in terms of $x$ and obtain
\begin{equation} \label{hren}
y'= H(J_2 +x),\quad H(J_2 +x) = [G']^{-1} (J_2 +x) .
\end{equation}
Substituting into the first equation, we obtain
\begin{equation} \label{lan}
y(x) =J_1 - G[H(J_2 +x)] + (J_2 +x)H(J_2 +x).
\end{equation}

\medskip

Now we are in a position to show how one can find 
a variational discrete model and its  conservation
laws 
by means of Lagrange-type technique. 
Let us choose a difference Lagrangian in the form
\begin{equation} \label{lan1}
{\cal L} = 
{ y + y_+ \over 2} + G ( y_x)   ,
\end{equation}
then 
$$
\begin{array}{l}
{\bf pr} X_{1} {\cal L} + {\cal L} { \dhp D } (\xi_{1} ) = 0 ; \\
\\
{\bf pr} X_{2} {\cal L} + {\cal L} { \dhp D } (\xi_{2} ) = 1= {
\dhp D } (x )  .\\
\end{array}
$$

The variations of ${\cal L}$ yield the following
quasiextremal equations:

\begin{equation} \label{ee21a}
{\displaystyle
{ \delta {\cal L} \over \delta y} : } \quad 
{\displaystyle   G'( y_x )  -   G'( y_{\bar{x}} )   
= { h_{+} + h_{-} \over 2 }    } ; \\
\end{equation}
\begin{equation} \label{ee21b}
{\displaystyle
{ \delta {\cal L} \over \delta x} :   } \quad 
{\displaystyle  - { y + y_+ \over 2 }   - G( y_x ) +
y_x G'( y_x )  
+ { y+ y_{-}  \over 2 } +  G( y_{\bar{x}} ) -
y_{\bar{x}} G'( y_{\bar{x}} ) 
  = 0     } . 
\end{equation}

Due to the invariance of the Lagrangian with respect
to the operators $X_{1}$
and $X_{2}$, the difference analog of 
Noether's theorem yields two first integrals

\begin{equation} \label{dd21}
I_{1} =  { y}   + G( y_x) - y_x G'( y_x) 
+ {x_+ - x \over 2 }  y_x,
\end{equation}

\begin{equation} \label{dot}
 I_{2}=  G'( y_x)  - { x + x_{+} \over 2} .
\end{equation}

As in the case of the algebra ${\bf D_{1,1} }$ we can solve for $y_x$
to obtain 
\begin{equation} \label{dura}
y_x = \Phi_1  (I_2 , { x + x_{+}} ).
\end{equation}

Substituting into the equation for $I_1$ we obtain
\begin{equation} \label{durak}
y = \Phi_2  (I_1, I_2 ,  x,  x_{+} ).
\end{equation}

Calculating $y_x$ from eq. (\ref{durak}) and setting it equal to (\ref{dura}),
we obtain a three point recursion relation for $x$. Solving it 
(if we can), we turn eq. (\ref{durak}) into an explicit general solution of
the difference scheme (\ref{ee21a}),(\ref{ee21b}).

\bigskip

\noindent {\bf Example~3.2}  Let us consider the case 
\begin{equation} \label{roma1}
{\cal L} = {{y + y_+}\over 2}  + \exp (  y_{x})  .
\end{equation}

The two first integrals (\ref{dd21}),(\ref{dot}) 
in this case are the following:
\begin{equation} \label{roma2}
I_1 = y +  \exp ( {y_{x}} )  - y_x \exp ( {y_x} )  + 
\frac{x_{n+1} - x_n}{ 2} y_x, 
\end{equation}
$$
\quad I_2 =  \exp ( {y_{x}} )  - \frac{x_{n+1} + x_n}{ 2}.
$$
Equation (\ref{dura}) and (\ref{durak}) reduce to 
\begin{equation} \label{roma3}
 y_x =  \ln \left( I_2 + \frac{x_{n+1} + x_n}{2} \right) ,
\end{equation}
\begin{equation} \label{roma4}
 y = I_1 - I_2 - {x_{n+1} + x_n  \over 2 } +
(I_2 + x_n) \ln \left( I_2 + \frac{x_{n+1} + x_n}{2}  \right),
\end{equation}
The recursion relation for $x$ is
\begin{equation} \label{roma5}
\frac{-x_{n+1} + x_{n-1}}{2}  + (I_2 + x_n)
[ \ln (2 I_2 +{x_{n+1} + x_n}) -\ln (2 I_2 +{x_{n} + x_{n-1}})].
\end{equation}
The last equation
is difficult to solve. We have however reduced a system of two
three-point equations to a single three-point one. 
We shall return to this case 
in Section 5 using an alternative method.

\bigskip

\noindent  ${ \bf D_{2,2} }$ The non--Abelian Lie
algebra
with non-connected elements 
\begin{equation} \label{op23}
X_1 = { \ddy} , \quad X_2 =  x { \ddx} + y  { \ddy}
\end{equation}
yields the invariant ODE
\begin{equation} \label{eq22}
y''= { 1 \over x} F(y').
\end{equation}

We define a function $G( y ')$   by the equation 
\begin{equation} \label{eq23}
F(y') = { G'( y ' ) \over G''( y ' ) }. 
\end{equation}
Then the  ODE (\ref{eq22}) is the Euler--Lagrangian equation for the
Lagrangian
\begin{equation} \label{lag23}
L = { 1 \over x } G ( y' )  ,
\end{equation}
which admits $X_{1}$ and $X_{2}$ as variational
symmetries:
$$
\begin{array}{l}
\mbox{pr} X_{1} L + L D( {\xi}_{1} ) = 0; \\
\\
\mbox{pr} X_{2} L + L D( {\xi}_{2} ) = 0  .\\
\end{array}
$$
Noether's theorem provides us with two 
first integrals: 
$$
J_{1}= { 1 \over x } G'( y') , \qquad 
J_{2}= G( y') + \left( { y \over x } - y' \right) G'(
y') . 
$$
 
\medskip

Let us take the difference Lagrangian
$$
{\cal L} = 
{ 2  \over x + x_{+} }  G ( y_x)   ,
$$
which satisfies 
$$
\begin{array}{l}
{\bf pr}  X_{1} {\cal L} + {\cal L} { \dhp D } (\xi_{1} ) = 0 ; 
\\
\\
{\bf pr}  X_{2} {\cal L} + {\cal L} { \dhp D } (\xi_{2} ) = 0  
. \\
\end{array}
$$
Then the variations of ${\cal L}$ yield the following 
quasiextremal equations:
\begin{equation} \label{eeq23}
\begin{array} {ll}
{\displaystyle
{ \delta {\cal L} \over \delta y} : } &
{\displaystyle  { 2 \over x + x_{+} } G' ( y_x) - 
{ 2 \over x_- + x } G' ( y_{\bar{x}} )    = 0   } ;
\\
\\
{\displaystyle
{ \delta {\cal L} \over \delta x} : } &
\begin{array} {l}
{\displaystyle  - { 2 h_{+} \over ( x + x_{+} )^2 } G(
y_x) 
+ { 2  \over ( x + x_{+} ) } G'( y_x) y_x 
- { 2  \over ( x + x_{+} ) } G( y_x) } \\
\\
{\displaystyle
- { 2 h_{-} \over ( x_-  + x )^2 } G( y_{\bar{x}} ) 
- { 2  \over ( x_-  + x ) } G'( y_{\bar{x}} )
y_{\bar{x}}  
+ { 2  \over ( x_- + x  ) } G( y_{\bar{x}} ) = 0 .  
     } 
\end{array} \\
\end{array}
\end{equation}

Since the Lagrangian is  invariant  
with respect to the operators $X_{1}$ and $X_{2}$, 
we find the first integrals
\begin{equation} \label{dd23}
{\displaystyle
I_{1} =  { 2    G' (y_x)   \over x + x_{+} }      , }
\qquad 
{\displaystyle
I_{2}=  {  4 x x_{+} \over ( x + x_{+}) ^2  } G( y_x) 
+ 
{ 2  G'(y_x)   \over {x + x_{+}} } 
\left( { y } - x y_x \right)  } 
\end{equation}
for the solutions of (\ref{eeq23}). 
\par
As in the case of the algebra ${ \bf D_{2,1}}$, 
we can solve for $y_x$, using the
integral $I_1$. We obtain:
\begin{equation} \label{her1}
y_x = \Phi_1 (I_1 , x + x_+).
\end{equation}
The second integral allows us to express $y$ as a function of
$x$ and $x_+$
\begin{equation} \label{her2}
y =x \Phi_1 + \frac{I_2}{I_1} - \frac{4x x_+}{I_1{(x + x_+)}^{2}}
G(\Phi_1).
\end{equation}

\medskip

\section{\large \bf Equations corresponding to Lagrangians
 invariant under three-dimensional Lie 
groups }

Among the ''prototype equations'' of our previous article [1], many 
have three-dimensional symmetry groups. In this section we shall 
consider two of these cases. Both of them come from Lagrangians 
that also have three-dimensional symmetry groups, i.e. 
all these symmetries are Lagrangian ones.

\medskip

\noindent  ${\bf D_{3,1} }$ Let us first consider a family of 
solvable Lie algebras depending on one constant $k$: 
\begin{equation} \label{op34c}
X_{1}= {\ddx} , \qquad  X_{2}= {\ddy} ,
\qquad  X_{3}= x {\ddx} + k y  {\ddy}, \qquad k \neq
0, { 1 \over 2 } ,  1, 2 . 
\end{equation}
The invariant equation has the form
\begin{equation} \label{eq34c}
y'' =  {y'} ^{ \textstyle { k - 2 \over k-1} } . 
\end{equation}

This equation can be obtained by the usual variational procedure from the 
Lagrangian

\begin{equation} \label{lag34c}
L = { (k-1) ^{2} \over k} ( y' )^ {\textstyle { k
\over k-1}}  +  y ,
\end{equation}

\noindent
which admits operators $X_{1}$ and $X_{2}$ 
for any parameter $k$ and  $X_{3}$ for $k = -1$: 
$$
\begin{array}{l}
\mbox{pr} X_{1} L +  L D( {\xi}_{1} ) = 0; \\
\\
\mbox{pr} X_{2} L +  L D( {\xi}_{2} ) = 1= D(x); \\
\\
\mbox{pr} X_{3} L +  L D( {\xi}_{3} ) = (k+1) L . \\
\end{array}
$$

It is possible to show that there is no Lagrangian
function
$L(x,y,y')$ which gives
eq.~(\ref{eq34c}) with  $k \neq -1$ as its Euler's
equation
and is divergence  invariant for 
all three symmetries (\ref{op34c}).

\medskip

For arbitrary $k$ there are two first integrals
$$
 J_{1} = { (1-k) \over k} ( y' )^ {\textstyle { k
\over k-1}} + y = A^0,
\qquad
J_{2}  =  (k-1) ( y' )^{\textstyle { 1 \over k-1} }  -x = B^0.
$$
Eliminating $y'$ we  find the general solution:
  \begin{equation} \label{sol34difc}
y = {1 \over k} \left( { 1 \over k-1 } \right)^{(k-1)}
( x + B^{0} ) ^k  +  A^{0}.
\end{equation}

\par
In the case $k = -1$ we have the further first integral 
corresponding to the symmetry $X_3$:
$$
 J_{3} = { 2 \over \sqrt { y'} } ( y - x y') +  xy = 
C^0.
$$
It is functionally dependent on $J_1$ and  $J_2$ since a second order 
ODE can possess only two functionally independent first integrals. 
Let us mention that the first integral $J_3$ 
is basic:

$$
J_1 = X_1 (J_3) , \qquad J_2 = - X_2 (J_3),
$$  
since
$$
[X_1 ,X_3] = X_1  , \qquad [ X_2 , X_3 ]  =  k X_2 .
$$
In this case we have the following relation:
\begin{equation} \label{relat}
4 - J_1 {J_2} - J_3 = 0.
\end{equation}
Thus, the integral $J_3$ is not 
independent and is of no use in the present context.

\medskip

\par

Now let us turn to the discrete case  and consider $ {\bf k=-1}$ only. Other 
values of $k$ will be considered in Section~5, using a different approach. 
Let us choose the Lagrangian to be
\begin{equation} \label{Lag11}
{\cal L} =   - 4
 \sqrt{ y_{x} }  +
{ y + y_+  \over 2} 
\end{equation}
as a discrete Lagrangian, which is invariant for $X_1$ and $X_3$
and divergence  invariant for $X_2$:
\begin{equation} \label{Lag12}
\begin{array}{l}
\mbox{pr} X_{1} {\cal L} + {\cal L} { \dhp D } (\xi_{1} ) = 0 ; 
\\
\mbox{pr} X_{2} {\cal L} + {\cal L} { \dhp D } (\xi_{2} ) = 1 =
{ \dhp D } (x) ;  
\\
\mbox{pr} X_{3} {\cal L} + {\cal L} { \dhp D } (\xi_{3} ) = 0. 
\end{array}
\end{equation} 

From the  Lagrangian  we obtain the 
quasiextremal equations:
\begin{equation} \label{eeq4c}
\begin{array} {ll}
{\displaystyle
{ \delta {\cal L} \over \delta y} : } &
{\displaystyle
-{ 4 \over h_{-} + h_{+} }
\left(
{ 1 \over \sqrt{y_{x}} }  
 -
{ 1 \over \sqrt{y_{\bar x}} } 
\right)
= 1  } ;  \\
\\
{\displaystyle
{ \delta {\cal L} \over \delta x} : } &
{\displaystyle
{ 4 }\left(
\sqrt{y_{x}} 
 -
\sqrt{y_{\bar x}} 
\right)
-  {y + y_+  \over 2} +  {y_-  + y  \over 2} =0. } \\
\end{array}
\end{equation}

\par
This system of equations is invariant with respect to
all three operators (\ref{op34c}). 
The application of the difference analog
 of the Noether theorem gives us 
 three first integrals:
\begin{equation} \label{dorc}
\begin{array}{c}
{ \displaystyle 
I_{1}= { -2 }
\sqrt{y_{x}}  +
 { y + y_+  \over 2 } = A , } \qquad 
{ \displaystyle 
I_{2}= -
{ 2 \over \sqrt{y_{x}} } -
 { x + x_+ \over 2 }  } = B,  \\
{ \displaystyle I_{3}= 
{ 2 ( x_{+} y  - y_{+} x ) \over h_{+} \sqrt{ y_{x} }
} + 
{  x_{+} y + y_{+} x \over 2  }   } = C   . 
\end{array}
\end{equation}

In contrast to the continuous case the three difference
first integrals 
$ I_{1}$, $ I_{2}$ and  $ I_{3}$  are functionally independent and instead 
of eq.~(\ref{relat})  
we have the following relation:
\begin{equation} \label{grid4d}
4 - I_1 I_2  - I_3 = { 1 \over 4} h_{+} ^2 y_{x} = 
\frac {4 \varepsilon^2 }{ (\varepsilon +2)^2  }. 
\end{equation}
This coincides with eq.~(\ref{relat}) in 
the continuous limit ${{\varepsilon}\to 0}.$
We see that 
the expression $h_{+} ^2 y_{x} $ is also a first
integral of (\ref{eeq4c}). This allows to introduce  
 a convenient lattice, namely:
\begin{equation} \label{grid34c}
 { 1 \over 4}  h_{-}  ^2 y_{\bar{x}} = 
 { 1 \over 4} h_{+}  ^2 y_{x} = 
\frac {4 \varepsilon^2 }{ (\varepsilon +2)^2  }   ,    
\quad \varepsilon =
const,   
\quad 0 <  \varepsilon \ll 1 . 
\end{equation}

Substituting $y_x$ from eq.~(\ref{grid34c}) into $I_2$, we obtain
 a two-term recursion relation for $x$, namely 
\begin{equation} \label{grid34h}
x_{n+1} - (1 + \varepsilon ) x_n - \varepsilon B = 0 ,
\end{equation}
or 
\begin{equation} \label{grida4h}
- (1 + \varepsilon ) x_{n+1} + x_n - \varepsilon B = 0 ,
\end{equation}
depending on the sign choice for $\sqrt{ y_{x}} $. 
These equations can be solved and we obtain a lattice satisfying 
\begin{equation} \label{grid34k}
x_{n} = ( x_0 + B )   (1 + \varepsilon )^n  - B , \quad x_0 > -B
\end{equation}
for the first equation and 
a lattice satisfying 
\begin{equation} \label{grida34k}
x_{n} = ( x_0 + B )   (1 + \varepsilon )^{-n}  - B , \quad x_0 < -B
\end{equation}
for the second equation. 
Using the expressions for $I_1$, we get the general solution for $y$ 
(the same for both lattices (\ref{grid34k}) and (\ref{grida34k})) as 
\begin{equation} \label{grid34m}
y_{n} = A - { 4 \over x_n + B } { 1 + \varepsilon   \over 
  (1 +  { \varepsilon \over 2 } )^2 }  .
\end{equation}
This agrees with the continuous case up to order $ \varepsilon $.

We have used the three integrals $I_1$,  $I_2$ and  $I_3$ 
to obtain the general   solution of the difference scheme 
(\ref{eeq4c}). Indeed, the solution (\ref{grid34k}),(\ref{grid34m}) 
for $x_n$, $y_n$ depends on 4 constants $(A,B,x_0,\varepsilon )$, 
as it should. 

The difference scheme is not compatible with a regular lattice, but 
requires an exponential one, as in eq.~(\ref{grid34k}). 
The only non-algebraic step in the integration was the solution of 
eq.~(\ref{grid34h}): a linear two point equation with constant 
coefficients.

\bigskip

\noindent  ${\bf D_{3,2} }$ The group given by the
operators
\begin{equation} \label{op32}
X_{1}= {\dx x} , \qquad  X_{2}= 2 x {\dx x} + y {\dx
y} ,
\qquad  X_{3}= x^{2} {\ddx} + xy {\ddy}
\end{equation}
corresponds to the invariant differential 
equation
\begin{equation} \label{eq3.14}
y'' =  y^{-3}.
\end{equation}
This equation can be obtained from the Lagrangian
function
\begin{equation} \label{lag32}
L =    {y'}^2   - { 1 \over  y ^2 },
\end{equation}
which admits all three operators:
\begin{equation} \label{lagf32}
\begin{array}{l}
\mbox{pr} X_{1} L +  L D( {\xi}_{1} ) = 0 ; \\
\\ 
\mbox{pr} X_{2} L  + L D( {\xi}_{2} ) = 0 ;  \\
\\
\mbox{pr} X_{3} L  + L D( {\xi}_{3} ) = 2 y'y = D ( y^2). \\
\end{array}
\end{equation}
Consequently, the symmetries yield the following
first integrals
\begin{equation} \label{lagh32}
\begin{array}{c}
{\displaystyle 
J_{1} =  {y'}^2   + { 1 \over  y ^2 } = A^0   , \qquad
J_{2} = 2  {x \over y^2} - 2(y - y'x) y' = 2B^0  ,}\\
\\
{\displaystyle 
J_{3} =  {x^2 \over y^2} + ( y - x y') ^2  = C ^0. } \\
\end{array}
\end{equation}

Using the integrals $A^0$ and $B^0$ we write the general solution $y(x)$ 
as 
\begin{equation} \label{sol3aa}
A^0 y^{2} = ( A^0 x - B^0 )^{2} + 1.
\end{equation}

We see that the third integral, denoted $J_3$  is not needed,  
is not useful and indeed, is not independent. The integrals 
$J_1$, $J_2$ and $J_3$ are related as follows:
\begin{equation} \label{solf3}
\left( { J_2 \over 2} \right) ^2 - J_1 J_3 + 1 = 0 . 
\end{equation}

\medskip

Now let us consider the discrete case. 
Let us consider the discrete Lagrangian function
\begin{equation} \label{lag32d}
{\cal L} =   y_{x}^2 - { 1 \over y y _+ }  ,
\end{equation}
which admits the same symmetries as the continuous one:
\begin{equation} \label{lagss32d}
\begin{array}{l}
{\bf pr} X_{1} {\cal L} +  {\cal L} { \dhp D } (\xi_{1} )= 0 ; 
\\
\\
{\bf pr} X_{2} {\cal L} +  {\cal L} { \dhp D } (\xi_{2} )= 0 ; 
\\
\\
{\bf pr} X_{3} {\cal L} +  {\cal L} { \dhp D } (\xi_{3} )
= \dd D(y^2).  \\
\end{array}
\end{equation}
The Lagrangian generates the invariant quasiextremal equations:
\begin{equation} \label{eeq314}
\begin{array} {ll}
{\displaystyle
{ \delta {\cal L} \over \delta y} : }&
{\displaystyle
2 (  y_{x} -  y_{\bar x} ) = { h_{+} \over y^2 y_{+} }
+
 { h_{-} \over y^2 y_{-} }
 }; \\
\\
{\displaystyle
{ \delta {\cal L} \over \delta x} : } &
{\displaystyle
( y_{x})^2 + { 1\over yy_{+} } -
( y_{\bar x})^2 - { 1 \over y y_{-} } =0
 }  . \\
\end{array}
\end{equation}
 
The  quasiextremal equations have three functionally
independent first integrals
\begin{equation} \label{int2}
\begin{array}{c} 
I_{1} =  {\displaystyle { y_x}^2 + \frac{1}{y y_{+} }} = A 
, 
\qquad 
I_{2} = {\displaystyle \frac{x + x_{+} }{ y y_{+} } +
2 { y_x} ( x_{+}  y_x - y_+  )   } = 2B , \\
\\
I_{3} = {\displaystyle   \frac{x x_{+}}{y y_{+} }  +
{ ( x_+ y_x  - y_+ )^2 } }  = C . \\
\end{array} 
\end{equation}

In the discrete case the integrals $I_1$, $I_2$ and $I_3$ are independent. 
Eq.~(\ref{solf3}) no longer holds and instead we have 
\begin{equation} \label{grid2}
\left( { I_2 \over 2 } \right) ^2 - I_1 I_3  + 1  = 
{ 1 \over 4 } \left( {h_{+} \over y y_{+} } \right)^2.
\end{equation}

In order to integrate the system (\ref{eeq314}) we will 
use three first integrals $A$, $B$ and the one in eq.~(\ref{grid2}), 
namely
\begin{equation} \label{gridss2}
 {h_{+} \over y y_{+} } = \varepsilon . 
\end{equation}
Eliminating $y_x$, $x_+$ and $y_+$,  we obtain the solution 
\begin{equation} \label{sol22}
Ay^2 = ( Ax - B ) ^2 + 1 - { \varepsilon ^2 \over 4 }  . 
\end{equation}
This agrees with the continuous limit (\ref{sol3aa}) up to order 
$ \varepsilon ^2 $. 

Calculating $y_x$ from eq.~(\ref{sol22}) and substituting into the expression 
for $A$ in eq.~(\ref{int2}) we obtain a two-point difference equation 
for $x$ and hence we obtain the lattice. In this case the equation has the form of a fractional linear mapping, i.e. it is a discrete version of the Riccati equation (with constant coefficients). 

Explicitly we obtain 
\begin{equation} \label{sdf1}
x_{n+1} = { \alpha x_{n} + \beta  \over \gamma x_n + \delta } 
\end{equation}

\begin{equation} \label{sdf2}
\begin{array}{lcllcl} 
\alpha & = & 1 -  \varepsilon B - { 1 \over 2 } \varepsilon^2  ;
& \beta  & = & {\displaystyle { \varepsilon \over A } } 
( 1 + B^2  - { 1 \over 4 } \varepsilon^2 ) ; \\
&&&&& \\
\gamma & = & - \varepsilon A ; 
& \delta & = & 1 + \varepsilon B - { 1 \over 2 } \varepsilon^2  .\\
\end {array}
\end{equation}
We see that the coefficients in the discrete Riccati equation
(\ref{sdf1})  satisfy
\begin{equation} \label{sdsss2}
  \alpha \delta - \beta \gamma = 1, \qquad 
  \alpha + \delta = 2 - \varepsilon^2.
\end{equation}
Like the continuous Riccati equation,
eq. (\ref{sdf1}) can be linearized. 
To do this we introduce a linear system 
\begin{equation} \label{sdf3}
\left( 
\begin{array}{c}
u_{n+1} \\
v_{n+1} \\
\end{array} 
\right)
= 
\left( 
\begin{array}{cc}
\alpha & \beta \\
\gamma & \delta  \\
\end{array} 
\right)
\left( 
\begin{array}{c}
u_{n} \\
v_{n} \\
\end{array} 
\right)
\end{equation}

If $u$ and $v$ satisfy eq.~(\ref{sdf3}), then 
\begin{equation} \label{griee2}
x = { u \over v} 
\end{equation}
will satisfy equation (\ref{sdf1}). Eq.~(\ref{sdf3}) can be 
solved by standard methods. Indeed, both $u$ and $v$ must satisfy 
\begin{equation} \label{griee3}
u_{n+2} - ( \alpha + \delta ) u_{n+1} 
+ ( \alpha \delta - \beta \gamma ) u_n = 0 .
\end{equation}

The characteristic equation 
\begin{equation} \label{griff2}
\lambda ^{2} - ( \alpha + \delta ) \lambda 
+ ( \alpha \delta - \beta \gamma )  = 0 
\end{equation}
is obtained by putting $ u_n = \lambda^n $.

In view of eq. (\ref{sdsss2}) the roots of eq. (\ref{griff2}) 
are complex.
The final result is that the solution of eq.~(\ref{sdf1}) is

\begin{equation} \label{grisss1}
x_n =  {  1 \over A } 
\sqrt{ 1 - { \varepsilon ^2 \over 4 } } 
\tan ( \omega n +\rho ) + { B \over A } , 
\end{equation}
where $\rho$ is an integration constant and $\omega$ satisfies
\begin{equation} \label{grsss2}
\tan \omega =  { 2 \varepsilon   \over 2 - \varepsilon^2 } 
\sqrt{ 1 - { \varepsilon ^2 \over 4 } } 
\end{equation}

Eqs. (\ref{sol22}) and (\ref{grisss1}) provide an explicit 
analytic solution of the
system (\ref{eeq314}). 
It is the general solution and involves 4 constants:
$A$, $B$, $\varepsilon$ and $\rho$. It follows from
eq.~(\ref{grisss1}) that the independent variable $x$ varies 
between $-$  and $+$ infinity as $\omega n + \rho$ varies between 
$\pm {\pi \over 2}$.

\bigskip

\section{\large \bf Integration of difference schemes with two
  variational symmetries  }

\subsection{ \bf  The method of perturbed Lagrangians }

We have mentioned in Section 3.2 that a two-dimensional group of
Lagrangian symmetries is always sufficient to reduce the original
system of two three-point equations to a single three-point equation
for the independent variable alone.

Here we shall show that in some cases we can do better. Using a
different approach, we will actually obtain a complete solution of a
difference scheme approximating a differential equation with a
Lagrangian, divergence invariant under a two-dimensional symmetry
group. 

The case we shall consider is the equation (\ref{eq21a}) and hence 
the two-dimensional Abelian group ${\bf D_{2,1} }$ corresponding to the
algebra (\ref{op21}). We shall make use of the fact that the
Lagrangian is not unique. Indeed we will consider three different
Lagrangians, all having the same continuous limit
(\ref{lag21}). Instead of writing the Lagrangian (\ref{lan1}) in the
discrete case, we shall use a family of Lagrangians, parametrized by
two constants, $\alpha$ and $\beta$: 
\begin{equation} \label{nn1}
{\cal L} = \alpha G(y_x) + \beta y + ( 1 - \beta ) y_+, \quad 
\alpha \approx 1, \quad 0 \leq \beta \leq 1. 
\end{equation}
Each Lagrangian provides its own quasiextremal system
\begin{equation} \label{nn2}
\alpha \left[ - G'(y_x) + G'(y_{\bar{x}}) \right] 
+ \beta h_+ + ( 1 - \beta) h_- = 0 
\end{equation}
\begin{equation} \label{nn3}
\alpha \left[ y_x G'(y_x) - y_{\bar{x}} G'(y_{\bar{x}}) 
-  G(y_x) + G(y_{\bar{x}}) \right] 
- \beta ( y - y_- )  - ( 1 - \beta) ( y_+ - y )  = 0 
\end{equation}
We shall view one Lagrangian, with $\alpha_3 = 1 $ and $\beta_3 = 0.5$
as the basic one, the other two as its perturbations. 

Each Lagrangian in the family is divergence invariant under 
$X_1 = \partial _x $ and $X_2 = \partial _y $ and hence provides two
first integrals of the corresponding quasiextremal equations
(\ref{nn2}) and (\ref{nn3}):
\begin{equation} \label{nn4}
\alpha \left[ - y_x G'(y_x) +  G(y_x)  \right] 
+ y + ( 1 - \beta) h_+ y_x   = A 
\end{equation}
\begin{equation} \label{nn5}
\alpha  G'(y_x)  
- x - \beta h_+    = B . 
\end{equation}

Let us now choose three different pairs $( \alpha _i , \beta _i )
$. They provide six integrals (and six quasiextremal equations). 
We shall show that by appropriately fine tuning the constants 
$\alpha _i$ and $ \beta _i $ and choosing some of the constants 
$A_i $ and $B_i$ we can manufacture a consistent difference system,
representing both the equation and the lattice. Moreover, we can
explicitly integrate the equations in a manner that approximates the
exact solution obtained in the continuous limit. 

Let us take one equation of the form (\ref{nn4}) and two of the form 
(\ref{nn5}). In these three equations we choose 
$\alpha_3 = 1 $, $\beta_3 = 0.5$ and $B_2 = B_3 = B $. We then take
the difference between the two equations involving $B$ to finally
obtain the following system of three two-point equations: 
\begin{equation} \label{nn6}
\alpha_1 \left[ - y_x G'(y_x) +  G(y_x)  \right] 
+ y + ( 1 - \beta_1 ) h_+ y_x   = A 
\end{equation}
\begin{equation} \label{nn7}
 G'(y_x)   - x - { 1 \over 2 } h_+    = B  
\end{equation}
\begin{equation} \label{nn8}
(1 - \alpha_2)   G'(y_x)  
- ( {1 \over 2 } - \beta_2  ) h_+    = 0 . 
\end{equation}

From eq.~(\ref{nn7}) and (\ref{nn8}) we have
\begin{equation} \label{nn9}
 G'(y_x)  =  { x_+ + x + 2B \over 2 }
\end{equation}
\begin{equation} \label{nn0}
x_+ - ( 1 + \varepsilon ) x - \varepsilon B = 0 , 
\end{equation}
where we have put 
\begin{equation} \label{pp1}
\varepsilon  = { 2 ( 1- \alpha_2) \over  \alpha_2 - 2 \beta _2 } . 
\end{equation}
The continuous limit will correspond to $ \varepsilon \rightarrow 0
$.

Eq.~(\ref{nn0}) coincides with eq.~(\ref{grid34h}) obtained using
three Lagrangian symmetries in a special case. Here it appears in a
much more general setting. The general solution of eq.~(\ref{nn0}) 
\begin{equation} \label{megh}
x_n = ( x_0 + B ) ( 1 + \varepsilon ) ^n - B 
\end{equation}
depends on one integration constant
$x_0$. This solution gives a lattice satisfying 
$h_- > 0$ and $h_+ > 0$ 
for $ x_0 > -  B$ if $ \varepsilon > 0 $ and 
for $ x_0 < -  B$ if $ \varepsilon < 0 $. 
For the other cases, namely $ x_0 < -  B$ if $ \varepsilon > 0 $ and 
for $ x_0 > -  B$ if $ \varepsilon < 0 $,  formula (\ref{megh}) gives a 
lattice with a reverse order of points:  $h_- < 0$ and $h_+ < 0$.

Using (\ref{megh}) and (\ref{nn9}), we can express $y_x$ in
terms of $x$. We have 
\begin{equation} \label{pp2}
 G'(y_x)  =  \left( 1 + { \varepsilon \over 2 } \right) ( B+ x) . 
\end{equation}
Denoting the inverse function of $ G'(y_x)$ as $H$,  we have 
\begin{equation} \label{pp3}
 y_x   =  H \left[  
\left( 1 + { \varepsilon \over 2 } \right) ( B+ x) 
\right]. 
\end{equation}
Using (\ref{nn6}) and (\ref{pp3}),  we can now write the general
solution of the system  (\ref{nn6}), (\ref{nn7}) and (\ref{nn8}) as 
\begin{equation} \label{pp4}
 y(x)    =  A - \alpha_1 G(H) + (x+ B ) H , 
\end{equation}
where we have put 
\begin{equation} \label{pp5}
\alpha_1 
\left( 1 + { \varepsilon \over 2 } \right) 
- ( 1 - \beta_1 )  \varepsilon = 1 . 
\end{equation}

The value of $ \alpha_1 $, still figuring in the solution (\ref{pp4}), 
must be so chosen as to obtain a consistent scheme. Indeed, $x_n$ and
$y_n$ given in eq.~(\ref{megh}) and (\ref{pp4}) will satisfy the
system   (\ref{nn6}), (\ref{nn7}) and (\ref{nn8}). We must however
assure that $y_x $ of eq.~(\ref{pp3}) and 
$y_x = ( y_{n+1} - y_n ) / ( x_{n+1} - x_n ) $ coincide. A simple
computation shows that this equality requires that $ \alpha_1 $ should
satisfy 
\begin{equation} \label{pp6}
\alpha_1 
= ( 1 + \varepsilon ) ^{n+1} ( x _0 + B ) 
{ H_{n+1 } - H_n \over G(H_{n+1 })  - G(H_n) }. 
\end{equation}
This equation is consistent only if the right hand side is a constant 
(independent on $n$). The constants $\alpha _i$ and $ \beta _i $ can 
depend upon the constant $ \varepsilon$ and for 
$  \varepsilon \rightarrow 0 $ we must have 
$  \alpha_1 ,  \alpha_2 \rightarrow  1 $; 
$  \beta_1 ,  \beta_2 \rightarrow  0.5 $.

From eq. (\ref{nn8}) we have
\begin{equation} \label{pddd1}
{ h_+ \over G'(y_x) } = { 2( 1- \alpha_2) \over 1-2 \beta_2 }  
\end{equation}
This expression must vanish for $\varepsilon \rightarrow  0$. 
To achieve this while respecting eq.~(\ref{pp1}) we put
\begin{equation} \label{pddd2}
\alpha_2 = 1 + \varepsilon ^2 , 
\qquad 
\beta_2 =  { 1 \over 2 } +  \varepsilon 
+ { \varepsilon ^2 \over 2  }. 
\end{equation}
Eq.~(\ref{pp1}) is satisfied exactly and we have
\begin{equation} \label{pddd3}
{ h_+ \over G'(y_x) } = { 2 \varepsilon  \over \varepsilon  + 2 } .   
\end{equation}
We can view  eq. (\ref{megh}) and (\ref{pp4}) 
as the general solution of the
following three point difference scheme:
\begin{equation} \label{pfff}
\begin{array}{l}
    {\displaystyle  G'(y_x) - G'(y_{\bar{x}} )  
- { x_+ - x_- \over 2 } = 0, }  \\
\\
    {\displaystyle   {  h_+  \over G'(y_x) }  
= { h_-  \over G'(y_{\bar{x}} ) }  .} \\
\end{array} 
\end{equation}

The system (\ref{pfff}) is invariant under the group corresponding to 
 ${\bf D_{2,1} } $. Strictly speaking, this is not a quasiextremal
system, 
since it can not be derived from any single Lagrangian. 
The arbitrary constants $A$, $B$ and $\varepsilon $ 
come from three first integrals~(\ref{nn6}), (\ref{nn7}) 
and (\ref{pddd3}) that are
associated with three different Lagrangians.

We have not proven that eq.~(\ref{pp6}) is consistent for arbitrary
functions $G(y_x)$. We shall however show below that in at least two
interesting special cases the above integration scheme is consistent. 

The results of this section can be summed up as a theorem.

\medskip
 
\noindent {\bf Theorem 5.1} The ODE~(\ref{eq21a}) obtained from the
Lagrangian~(\ref{lag21}) can be approximated by the difference 
system~(\ref{pfff}). If $\alpha_1$ of
eq.~(\ref{pp6}) is constant, then the general solution of this system
is given by 
\begin{equation} \label{pp7}
\begin{array}{l}
x_n = ( x_0  + B ) ( 1 +  \varepsilon ) ^{n} - B ,  \\
y(x_n )  =  A - \alpha_1 G(H_n ) + (x_n + B ) H_n , \\
\end{array} 
\end{equation}
where $ A$, $B$, $\varepsilon$ and $x_0$ are arbitrary constants. 
For  $\varepsilon \rightarrow 0 $,  $y(x_n)$ agrees with the solution 
(\ref{lan}) of the ODE~(\ref{eq21a}).

\medskip

As applications of this theorem let us consider two different
equations, each invariant under a three-dimensional group with 
${\bf D_{2,1} }$ as an invariant subgroup. In both cases the Lagrangian is
only divergence invariant under the subgroup ${\bf D_{2,1} }$.

\bigskip

\subsection{ \bf A polynomial nonlinearity. }

\begin{equation} \label{vv1}
{\bf D_{3,1} }: \qquad X_{1}= {\ddx} , \quad  X_{2}= {\ddy} ,
\quad  X_{3}= x {\ddx} + k y  {\ddy}, \quad k \neq
0, { 1 \over 2 } ,  \pm 1, 2 
\end{equation}

This algebra for $ k = -1 $ was treated in Section~4, now we consider
the generic  case. We take 
\begin{equation} \label{vv2}
G( y_x) = { (k-1)^2 \over k } {y_x} ^{ \textstyle { k \over k-1} } 
\end{equation}
and hence 
\begin{equation} \label{vv3}
G' ( y_x) = (k-1) {y_x} ^{ \textstyle { 1 \over k-1} } 
= \left( 1 + {\varepsilon \over 2 } \right) ( x + B ) .
\end{equation}
Eq.~(\ref{pp3}) reduces to  
\begin{equation} \label{vv4}
y_x   =  H_n (x) 
= \left( { x + B \over k-1} \right) ^{k-1}  
\left( 1 + {\varepsilon \over 2 } \right) ^{k-1}  
\end{equation}
and we have
\begin{equation} \label{vv5}
G(H_n) = { (k-1)^2 \over k }  
\left( { x + B \over k-1} \right) ^{k}  
\left( 1 + {\varepsilon \over 2 } \right) ^{k}  .
\end{equation}
Substituting into (\ref{pp6}), we find 
\begin{equation} \label{vv6}
\alpha _1 = 
{ k   ( 1 + \varepsilon )  ( ( 1 + \varepsilon) ^{k-1} - 1)  
\over 
(k  -1) ( 1 + {\varepsilon \over 2 } ) ( ( 1 + \varepsilon ) ^k - 1)   }
\end{equation}
so that we have $ \alpha _1  = 1 + O ( \varepsilon ^2 ) $.

Thus, $ \alpha _1$ is a constant, close to $ \alpha _1  = 1 $ for 
$  \varepsilon \ll 1$. The solution $y_n$ of (\ref{pp7}) specializes to 
\begin{equation} \label{vv7}
y_n = A + { (x + B)^k  \over (k-1) ^{k-1} }
{  \varepsilon  \left( 1 + {\varepsilon \over 2 } \right) ^{k-1} 
\over ( 1 + \varepsilon ) ^k - 1  }. 
\end{equation}
This agrees with the solution (\ref{sol34difc}) of the ODE 
(\ref{eq34c}) up to  $ O( \varepsilon ^2) $.

It is interesting to note that for $k=-1$ $ \alpha_1$ becomes
  independent on $ \varepsilon$ and we obtain $ \alpha_1 = 1 $,
  $\beta_1 = 0.5$. The solution (\ref{vv7}) provides us with the
  solution (\ref{grid34m}), which was obtained in Section 4 with the
  help of a different method.

\bigskip

\subsection{ \bf An exponential nonlinearity }

We consider another three-dimensional group and its Lie algebra,
namely: 
\begin{equation} \label{ww1}
{\bf D_{3,3} }: \qquad  X_{1}= {\ddx} , \qquad  X_{2}= {\ddy},
\qquad  X_{3}= x {\ddx} + (x+y ) {\ddy}. 
\end{equation}
The corresponding invariant ODE is 
\begin{equation} \label{ww2}
y'' =  \exp ( -y' )
\end{equation}
and can be obtained from the Lagrangian
\begin{equation} \label{ww3}
L = \exp( y' ) + y.
\end{equation}

We have 
\begin{equation} \label{ww4}
\begin{array}{l}
\mbox{pr} X_{1} L + L D( {\xi}_{1} ) = 0; \\
\mbox{pr} X_{2} L + L D( {\xi}_{2} ) = 1 = D(x) ;\\
\end{array}
\end{equation}
The corresponding first integrals of eq.~(\ref{ww2}) are 
\begin{equation} \label{ww5}
  \exp ( y' ) ( 1 - y' ) + y = A , \qquad 
 \exp ( y' )  - x  = B .
\end{equation}
Finally, the general solution of eq.~(\ref{ww2}) is 
\begin{equation}    \label{ww6} 
y = (x +  B) ( \ln ( x+ B ) - 1 ) + A.
\end{equation}

Now let us consider the discrete case, following the method of
Section~5.1. We have 
\begin{equation}    \label{ww7} 
G( y_x ) =  \exp ( y_x ) 
\end{equation}
and hence 
\begin{equation}    \label{ww8} 
\begin{array}{l}
G'(y_x) =  \exp ( y_x ) = ( x_n + B ) 
\left( 1 + {\varepsilon \over 2 } \right) \\
H_n = y_x = \ln ( x_n + B ) 
+ \ln  \left( 1 + {\varepsilon \over 2 } \right) \\
\end{array}
\end{equation}
Substituting into  eq.~(\ref{pp6}), we find 
\begin{equation}    \label{ww9} 
\alpha_1 = { ( 1 + \varepsilon) \ln ( 1 + \varepsilon) 
\over \varepsilon \left( 1 + {\varepsilon \over 2 } \right) }
\end{equation}
so that $ \alpha_1$ is indeed a constant and moreover we have 
$ \alpha _1  = 1 + O ( \varepsilon ^2 ) $. 

The solution $y(x) $ on the lattice given in eq.~(\ref{pp7}) is 
\begin{equation}    \label{ww0} 
y_n = A +  (x_n  +  B) \ln ( x_n + B ) 
+ (x_n + B ) 
\left[ 
\ln \left( 1 + { \varepsilon \over 2}  \right) 
- { (1 +  \varepsilon ) \ln  ( 1 +  \varepsilon ) \over \varepsilon }  
\right] .
\end{equation}

This agrees with the solution (\ref{ww6}) of the ODE (\ref{ww2})  
up to $O(  \varepsilon ^2) $.

\medskip

\section{\large \bf Concluding remarks}

We see that variational symmetries, and the first integrals they
provide, 
play a crucial role in the study of exact solutions of invariant
difference schemes. Much more so than in the theory of ordinary
differential equations.

  The procedure that we followed in this article can be reformulated as
follows. We start from the continuous case where we know a Lagrangian
density $L(x,y,y')$, invariant under a group $G_0$ of local point
transformations, i.e. satisfying 
condition (\ref{cond}), or (\ref{cong}). 
We hence also know the corresponding Euler-Lagrange
equation, invariant under the same group, or a larger group containing 
$G_0$
as a subgroup.

  Let us assume that we can approximate this Lagrangian by a "discrete 
Lagrangian density" $L(x,y, x_+,y_+)$ invariant under the same group $G_0$.
Even in the absence of any symmetry group, the Lagrangian will provide 
us with the quasiextremal equations (\ref{d55}), i.e. with a discrete 
Euler-Lagrange system. This system can be identified with the
difference system (\ref{sysa1}), (\ref{sysa2}).

If the Lagrangian is invariant under a one-dimensional symmetry group,
we can reduce the quasiextremal system to a three-point relation for x
alone, plus a "discrete quadrature" for y (see Section~3.1). If the
symmetry group of the Lagrangian is two-dimensional, we can always reduce
the quasiextremal system to one three-point equation for x alone, and 
write the solution $y_n(x)$ directly (see Section~3.2).

If the invariance group of the Lagrangian is (at least)
three-dimensional then we can integrate the system explicitly (Section~4).

Finally,  we have shown that if the symmetry group of the Lagrangian
  is two-dimensional, but the quasiextremal system has a third 
(non-Lagrangian) symmetry, we can also integrate explicitly.

\medskip
\bigskip

\noindent{\bf \large Acknowledgments}

\medskip

While working on this project, we benefited from a NATO collaborative
research grant  PST.CLG.978431, which made visits of V.D. to the CRM 
possible. The research of V.D. was sponsored 
in part by the Russian Fund for Basic Research 
under research project No 03-01-00446. 
The research of
R.K. was supported by the Norwegian Research
Council under contracts no.111038/410, through the
SYNODE project, and no.135420/431, through
the BeMatA program. 
The research of P.W. was partly supported by 
research grants from NSERC of Canada and FQRNT du Qu{\'e}bec.

\medskip
\bigskip


\end{document}